\documentstyle[preprint,eqsecnum,aps]{revtex}

\begin{document}
\draft
\preprint{number}
\title{The Behavior of Kasner Cosmologies with Induced Matter}
\author{Paul Halpern}
\address{Department of Mathematics, Physics and Computer Science\\
University of the Sciences in Philadelphia\\
600 S. 43rd Street\\
Philadelphia, Pa. 19104}
\date{May 10, 2000}
\maketitle
\begin{abstract}
We extend the induced matter model, previously applied to a variety of isotropic cases,
to a generalization of Bianchi type-I anisotropic cosmologies. The induced matter model is a 5D Kaluza-Klein approach in which assumptions of compactness are relaxed for the fifth coordinate, leading to extra geometric terms.  One interpretation of these extra terms is to identify them as an ``induced matter'' contribution to the stress-energy tensor.    
In similar spirit, we construct a five dimensional metric in which the spatial slices possess Bianchi type-I geometry.  We find a set of solutions for the five dimensional Einstein equations, and determine the pressure and density of induced matter. We comment on the long-term dynamics of the model, showing that the assumption of positive density leads to the contraction over time of the fifth scale factor.
\end{abstract}
\pacs{04.50.+h, 98.80.Cq, 98.80.Hw}

\narrowtext

\section{Introduction}
\label{sec:intro}

An intriguing notion suggested by Einstein is that the properties of matter in general relativity might have purely geometric origin \cite{einstein}.  Paul S. Wesson has recently revived this idea within the context of five dimensional Kaluza-Klein cosmologies \cite{wesson1}. 
In standard Kaluza-Klein models, extra dimensions are required to be compact, and hence unobservable at present.  Following the idea that cosmological dynamics might naturally reduce the size of higher dimensions \cite{chodos,freund,demianski,halpern}, Wesson relaxes this assumption of compactness. Wesson further postulates that the impact of the fifth dimension might be felt in the four dimensional universe through the presence of ``induced matter.''  By moving all terms related to the fifth dimension from the geometric to the energy-momentum side of the vacuum field equations, he naturally introduces a way of describing matter geometrically.  Thus not just electromagnetic phenemona, but also material properties such as densities and pressures, are molded by the extra dimension \cite{overduin}.

Wesson considers the 5D extension of the flat 4D metric.  He derives the Einstein tensor for this metric and sets it equal to zero.  He then collects all the terms in $G^0_0$ dependent on the fifth scale factor or on derivatives with respect to the fifth coordinate, and identifies the density of the induced matter with this total.  Similarly, he collects all the terms in $G^1_1$ ($ = G^2_2$ = $G^3_3$) that depend on the fifth scale factor or on derivatives of the fifth coordinate and identifies the pressure of the induced matter with this sum.  Remarkably, he finds that the properties of this matter obey the same relationships as in the standard four-dimensional case.  Namely, he shows that the perfect fluid described by the induced matter density and pressure satisfies both the first law of thermodynamics and Newton's law of motion.  Furthermore, he demonstrates that one particular solutions of the Einstein equations for his metric satisfies the equation of state for radiation, and another demonstrates the type of behavior associated with Einstein-de Sitter cosmologies \cite{wesson2}. 

These results pertain to the special situation of spatially isotropic geometries.  However, it is instructive to examine the more general case of anisotropic models.  These models are particularly of interest when examining the properties of the very early universe.  As Belinskii, Khalatnikov and Lifshitz demonstrated \cite{bkl}, the behavior of the most general approach to the initial singularity (looking backward in time) might be well modeled by anisotropic (Bianchi-type) cosmologies.  As the chaotic cosmology program suggests, close to the initial singularity, conditions may have been far less regular than they are today \cite{barrow}.  Perhaps, as some authors have suggested, the same mechanism underlying cosmological dimensional reduction might have also led to isotropization? \cite{biesada,demaret}

To this aim, we have extended the Bianchi anisotropic geometries by adding an extra dimension.  We hope to generalize Wesson's results by finding and analyzing solutions of the Einstein equations for these models, and by examining the density and pressure of the associated induced matter.  We have begun our study by looking at Bianchi type-I;  future studies will focus on other Bianchi types.

\section{The Generalized Bianchi type-I Model}
\label{sec:model}

We construct the 5D extension of the 4D Bianchi type-I metric:

\begin{equation}
ds^2 = e^{\nu} dt^2 - e^{\alpha} dx^2 - e^{\beta} dy^2 - e^{\gamma}dz^2 - e^{\mu} d{\psi}^2
\end{equation}

Here, following Wesson's notation, we supplement the time coordinate $t$ and the three spatial coordinates $x$, $y$ and $z$ with a fifth coordinate $\psi$.  We assume that the metric cooefficients $\mu$, $\nu$, $\alpha$, $\beta$ and $\gamma$ each depend, in general, on both $t$ and $\psi$.  (An earlier study by Socorro, Villanueva and Pimental \cite{socorro} considered a similar model, but assumed only time dependence of the scale factors.) We use overdots to represent partial derivatives with respect to $t$, and asterisks to represent partial derivatives with respect to $\psi$.

We find the non-zero components of the Einstein tensor for this metric:

\begin{eqnarray}
\label{eqn:ein1}
G^0_0= e^{-\nu} ( & & -\frac{1}{4}{{\mu}^.}{{\alpha}^.}-\frac{1}{4}{{\mu}^.}{{\beta}^.}-\frac{1}{4}{{\mu}^.}{{\gamma}^.}-\frac{1}{4}{{\alpha}^.}{{\beta}^.}-\frac{1}{4}{{\alpha}^.}{{\gamma}^.}-\frac{1}{4}{{\beta}^.}{{\gamma}^.})+ \nonumber \\ e^{-\mu} ( & &\frac{1}{2} {\alpha}^{**} +\frac{1}{2} {\beta}^{**}+ \frac{1}{2} {\gamma}^{**} + \frac{1}{4} {{\alpha}^*}^2 + \frac{1}{4} {{\beta}^*}^2 + \frac{1}{4} {{\gamma}^*}^2- \nonumber \\& & \frac{1}{4}{{\mu}^*}{{\alpha}^*}-\frac{1}{4}{{\mu}^*}{{\beta}^*}-\frac{1}{4}{{\mu}^*}{{\gamma}^*}+\frac{1}{4}{{\alpha}^*}{{\beta}^*}+\frac{1}{4}{{\alpha}^*}{{\gamma}^*}+\frac{1}{4}{{\beta}^*}{{\gamma}^*}) \\ \nonumber \\ 
\label{eqn:ein2}
G^0_4= e^{-\nu} ( & & 2{\alpha^{.*}} + 2{\beta^{.*}} + 2{\gamma^{.*}} +{{\alpha}^.}{{\alpha}^*} + {{\beta}^.}{{\beta}^*} + {{\gamma}^.}{{\gamma}^*} - \nonumber \\ & & {{\alpha}^.}{{\nu}^*}- {{\beta}^.}{{\nu}^*}- {{\gamma}^.}{{\nu}^*} - {{\alpha}^*}{{\mu}^.}- {{\beta}^*}{{\mu}^.}- {{\gamma}^*}{{\mu}^.})\\ \nonumber \\ 
\label{eqn:ein3}
G^1_1=   e^{-\nu} ( & & -\frac{1}{2}{{\beta}^{..}}-\frac{1}{2}{{\gamma}^{..}}-\frac{1}{2}{{\mu}^{..}}-\frac{1}{4}{{\beta}^.}^2-\frac{1}{4}{{\gamma}^.}^2-\frac{1}{4}{{\mu}^.}^2- \nonumber \\ & & \frac{1}{4}{{\beta}^.}{{\gamma}^.} -\frac{1}{4}{{\beta}^.}{{\mu}^.} - \frac{1}{4}{{\gamma}^.}{{\mu}^.} + \frac{1}{4}{{\beta}^.}{{\nu}^.} +\frac{1}{4}{{\gamma}^.}{{\nu}^.} + \frac{1}{4}{{\mu}^.}{{\nu}^.}) + \nonumber \\ e^{-\mu} ( & & \frac{1}{2}{{\beta}^{**}} + \frac{1}{2}{{\gamma}^{**}} + \frac{1}{2}{{\nu}^{**}} + \frac{1}{4}{{\beta}^*}^2 +\frac{1}{4}{{\gamma}^*}^2 + \frac{1}{4}{{\nu}^*}^2+ \nonumber \\ & & \frac{1}{4} {{\beta}^*}{{\gamma}^*} + \frac{1}{4} {{\beta}^*}{{\nu}^*} +\frac{1}{4} {{\gamma}^*}{{\nu}^*} - \frac{1}{4}{{\beta}^*}{{\mu}^*} - \frac{1}{4}{{\gamma}^*}{{\mu}^*} - \frac{1}{4}{{\mu}^*}{{\nu}^*}) \\ \nonumber \\
\label{eqn:ein4}
G^2_2=   e^{-\nu} ( & & -\frac{1}{2}{{\alpha}^{..}}-\frac{1}{2}{{\gamma}^{..}}-\frac{1}{2}{{\mu}^{..}}-\frac{1}{4}{{\alpha}^.}^2-\frac{1}{4}{{\gamma}^.}^2-\frac{1}{4}{{\mu}^.}^2- \nonumber \\ & & \frac{1}{4}{{\alpha}^.}{{\gamma}^.} -\frac{1}{4}{{\alpha}^.}{{\mu}^.} - \frac{1}{4}{{\gamma}^.}{{\mu}^.} + \frac{1}{4}{{\alpha}^.}{{\nu}^.} +\frac{1}{4}{{\gamma}^.}{{\nu}^.} + \frac{1}{4}{{\mu}^.}{{\nu}^.}) + \nonumber \\ e^{-\mu} ( & & \frac{1}{2}{{\alpha}^{**}} + \frac{1}{2}{{\gamma}^{**}} + \frac{1}{2}{{\nu}^{**}} + \frac{1}{4}{{\alpha}^*}^2 +\frac{1}{4}{{\gamma}^*}^2 + \frac{1}{4}{{\nu}^*}^2+ \nonumber \\ & & \frac{1}{4} {{\alpha}^*}{{\gamma}^*} + \frac{1}{4} {{\alpha}^*}{{\nu}^*} +\frac{1}{4} {{\gamma}^*}{{\nu}^*} - \frac{1}{4}{{\alpha}^*}{{\mu}^*} - \frac{1}{4}{{\gamma}^*}{{\mu}^*} - \frac{1}{4}{{\mu}^*}{{\nu}^*}) \\ \nonumber \\
\label{eqn:ein5}
G^3_3=   e^{-\nu} ( & & -\frac{1}{2}{{\alpha}^{..}}-\frac{1}{2}{{\beta}^{..}}-\frac{1}{2}{{\mu}^{..}}-\frac{1}{4}{{\alpha}^.}^2-\frac{1}{4}{{\beta}^.}^2-\frac{1}{4}{{\mu}^.}^2- \nonumber \\ & & \frac{1}{4}{{\alpha}^.}{{\beta}^.} -\frac{1}{4}{{\alpha}^.}{{\mu}^.} - \frac{1}{4}{{\beta}^.}{{\mu}^.} + \frac{1}{4}{{\alpha}^.}{{\nu}^.} +\frac{1}{4}{{\beta}^.}{{\nu}^.} + \frac{1}{4}{{\mu}^.}{{\nu}^.}) + \nonumber \\ e^{-\mu} ( & & \frac{1}{2}{{\alpha}^{**}} + \frac{1}{2}{{\beta}^{**}} + \frac{1}{2}{{\nu}^{**}} + \frac{1}{4}{{\alpha}^*}^2 +\frac{1}{4}{{\beta}^*}^2 + \frac{1}{4}{{\nu}^*}^2+ \nonumber \\ & & \frac{1}{4} {{\alpha}^*}{{\beta}^*} + \frac{1}{4} {{\alpha}^*}{{\nu}^*} +\frac{1}{4} {{\beta}^*}{{\nu}^*} - \frac{1}{4}{{\alpha}^*}{{\mu}^*} + \frac{1}{4}{{\beta}^*}{{\mu}^*} - \frac{1}{4}{{\mu}^*}{{\nu}^*}) \\ \nonumber \\
\label{eqn:ein6}
G^4_4=   e^{-\nu} ( & & -\frac{1}{2}{{\alpha}^{..}} -\frac{1}{2}{{\beta}^{..}}-\frac{1}{2}{{\gamma}^{..}} -\frac{1}{4}{{\alpha}^.}^2-\frac{1}{4}{{\beta}^.}^2-\frac{1}{4}{{\gamma}^.}^2- \nonumber \\ & & \frac{1}{4}{{\alpha}^.}{{\beta}^.} -\frac{1}{4}{{\alpha}^.}{{\gamma}^.} - \frac{1}{4}{{\beta}^.}{{\gamma}^.} + \frac{1}{4}{{\alpha}^.}{{\nu}^.} +\frac{1}{4}{{\beta}^.}{{\nu}^.} + \frac{1}{4}{{\gamma}^.}{{\nu}^.}) + \nonumber \\ e^{-\mu} ( & & \frac{1}{4} {{\alpha}^*}{{\beta}^*} + \frac{1}{4} {{\alpha}^*}{{\gamma}^*} +\frac{1}{4} {{\beta}^*}{{\gamma}^*} +\frac{1}{4} {{\alpha}^*}{{\nu}^*} + \frac{1}{4} {{\beta}^*}{{\nu}^*} + \frac{1}{4}{{\gamma}^*}{{\nu}^*})
\end{eqnarray}

In looking at the 5D ``vacuum'' case, we set each of these Einstein tensor components (eqns. \ref{eqn:ein1}- \ref{eqn:ein6}) equal to zero.  Then, we collect each of the terms in $G^0_0$ dependent on either $\mu$ or on derivatives with respect to $\psi$ and identify this quantity with the 4D induced matter density:

\begin{eqnarray}
\label{eqn:density}
\rho= e^{-\nu} ( & & -\frac{1}{4} {{\alpha}^.}{{\mu}^.}- \frac{1}{4} {{\beta}^.}{{\mu}^.}- \frac{1}{4} {{\gamma}^.}{{\mu}^.}) + \nonumber \\ e^{-\mu} ( & & \frac{1}{2} {\alpha}^{**} +\frac{1}{2} {\beta}^{**}+ \frac{1}{2} {\gamma}^{**} + \frac{1}{4} {{\alpha}^*}^2 + \frac{1}{4} {{\beta}^*}^2 + \frac{1}{4} {{\gamma}^*}^2- \nonumber \\ & & \frac{1}{4}{{\mu}^*}{{\alpha}^*}-\frac{1}{4}{{\mu}^*}{{\beta}^*}-\frac{1}{4}{{\mu}^*}{{\gamma}^*}+\frac{1}{4}{{\alpha}^*}{{\beta}^*}+\frac{1}{4}{{\alpha}^*}{{\gamma}^*}+\frac{1}{4}{{\beta}^*}{{\gamma}^*})
\end{eqnarray} 

Similarly, we collect each of the terms in $G^1_1$, $G^2_2$ and $G^3_3$ dependent on either $\mu$ or on derivatives in $\psi$ and identify each sum with the respective components of the 4D induced matter pressure:

\begin{eqnarray}
\label{eqn:press1}
P_1 = e^{-\nu} ( & & \frac{1}{2}{{\mu}^{..}}+ \frac{1}{4}{{\mu}^.}^2+ \frac{1}{4} {{\beta}^.}{{\mu}^.}+ \frac{1}{4} {{\gamma}^.}{{\mu}^.}+ \frac{1}{4} {{\mu}^.}{{\nu}^.}) + \nonumber \\ e^{-\mu} ( & & -\frac{1}{2}{{\beta}^{**}} - \frac{1}{2}{{\gamma}^{**}} -\frac{1}{2}{{\nu}^{**}} - \frac{1}{4}{{\beta}^*}^2 -\frac{1}{4}{{\gamma}^*}^2 - \frac{1}{4}{{\nu}^*}^2- \nonumber \\ & & \frac{1}{4} {{\beta}^*}{{\gamma}^*} - \frac{1}{4} {{\beta}^*}{{\nu}^*} -\frac{1}{4} {{\gamma}^*}{{\nu}^*} + \frac{1}{4}{{\beta}^*}{{\mu}^*} + \frac{1}{4}{{\gamma}^*}{{\mu}^*} + \frac{1}{4}{{\mu}^*}{{\nu}^*})\\ \nonumber \\
\label{eqn:press2}
P_2 = e^{-\nu} ( & & \frac{1}{2}{{\mu}^{..}}+ \frac{1}{4}{{\mu}^.}^2+ \frac{1}{4} {{\alpha}^.}{{\mu}^.}+ \frac{1}{4} {{\gamma}^.}{{\mu}^.}+ \frac{1}{4} {{\mu}^.}{{\nu}^.}) +\nonumber \\ e^{-\mu} ( & & -\frac{1}{2}{{\alpha}^{**}} - \frac{1}{2}{{\gamma}^{**}} - \frac{1}{2}{{\nu}^{**}} + \frac{1}{4}{{\alpha}^*}^2 +\frac{1}{4}{{\gamma}^*}^2 + \frac{1}{4}{{\nu}^*}^2- \nonumber \\ & & \frac{1}{4} {{\alpha}^*}{{\gamma}^*} - \frac{1}{4} {{\alpha}^*}{{\nu}^*} -\frac{1}{4} {{\gamma}^*}{{\nu}^*} + \frac{1}{4}{{\alpha}^*}{{\mu}^*} + \frac{1}{4}{{\gamma}^*}{{\mu}^*} + \frac{1}{4}{{\mu}^*}{{\nu}^*})\\ \nonumber \\
\label{eqn:press3}
P_3 = e^{-\nu} ( & & \frac{1}{2}{{\mu}^{..}}+ \frac{1}{4}{{\mu}^.}^2+ \frac{1}{4} {{\alpha}^.}{{\mu}^.}+ \frac{1}{4} {{\beta}^.}{{\mu}^.}+ \frac{1}{4} {{\mu}^.}{{\nu}^.}) + \nonumber \\ e^{-\mu} ( & & -\frac{1}{2}{{\alpha}^{**}} - \frac{1}{2}{{\beta}^{**}} - \frac{1}{2}{{\nu}^{**}} - \frac{1}{4}{{\alpha}^*}^2 -\frac{1}{4}{{\beta}^*}^2 - \frac{1}{4}{{\nu}^*}^2- \nonumber \\ & & \frac{1}{4} {{\alpha}^*}{{\beta}^*} - \frac{1}{4} {{\alpha}^*}{{\nu}^*} -\frac{1}{4} {{\beta}^*}{{\nu}^*} + \frac{1}{4}{{\alpha}^*}{{\mu}^*} + \frac{1}{4}{{\beta}^*}{{\mu}^*} + \frac{1}{4}{{\mu}^*}{{\nu}^*})
\end{eqnarray}
Writing the Einstein equations in terms of these density and pressure components, the 5D components (\ref{eqn:ein1}- \ref{eqn:ein6}) reduce to the equivalent of 4D form:
\begin{eqnarray}
\label{eqn:einstein0}
e^{-\nu} ( & & -\frac{1}{4}{{\alpha}^.}{{\beta}^.}-\frac{1}{4}{{\alpha}^.}{{\gamma}^.}-\frac{1}{4}{{\beta}^.}{{\gamma}^.})+ \rho = 0\\ \nonumber \\ 
\label{eqn:einstein1}
e^{-\nu} ( & & -\frac{1}{2}{{\beta}^{..}}-\frac{1}{2}{{\gamma}^{..}}-\frac{1}{4}{{\beta}^.}^2-\frac{1}{4}{{\gamma}^.}^2- \frac{1}{4}{{\beta}^.}{{\gamma}^.} + \frac{1}{4}{{\beta}^.}{{\nu}^.} +\frac{1}{4}{{\gamma}^.}{{\nu}^.}) + P_1 = 0 \\ \nonumber \\
\label{eqn:einstein2}
e^{-\nu} ( & & -\frac{1}{2}{{\alpha}^{..}}-\frac{1}{2}{{\gamma}^{..}}- \frac{1}{4}{{\alpha}^.}^2-\frac{1}{4}{{\gamma}^.}^2- \frac{1}{4}{{\alpha}^.}{{\gamma}^.}  + \frac{1}{4}{{\alpha}^.}{{\nu}^.} +\frac{1}{4}{{\gamma}^.}{{\nu}^.}) + P_2 = 0 \\ \nonumber \\ 
\label{eqn:einstein3}
e^{-\nu} ( & & -\frac{1}{2}{{\alpha}^{..}}-\frac{1}{2}{{\beta}^{..}}-\frac{1}{4}{{\alpha}^.}^2-\frac{1}{4}{{\beta}^.}^2- \frac{1}{4}{{\alpha}^.}{{\beta}^.} + \frac{1}{4}{{\alpha}^.}{{\nu}^.} +\frac{1}{4}{{\beta}^.}{{\nu}^.}) + P_3 = 0
\end{eqnarray}

Generalizing for this Bianchi type-I model the relationships found by Wesson \cite{wesson2} for the isotropic case, these equations determine the behavior of the induced matter, governing its material properties.  Since, in general, the three pressure terms are unequal, we cannot consider the induced matter a perfect fluid.

Two additional equations further govern the model's behavior: 
\begin{eqnarray}
\label{eqn: einstein4}
e^{-\nu} ( & & 2{\alpha^{.*}} + 2{\beta^{.*}} + 2{\gamma^{.*}} +{{\alpha}^.}{{\alpha}^*} + {{\beta}^.}{{\beta}^*} + {{\gamma}^.}{{\gamma}^*} - \nonumber \\ & & {{\alpha}^.}{{\nu}^*}- {{\beta}^.}{{\nu}^*}- {{\gamma}^.}{{\nu}^*} - {{\alpha}^*}{{\mu}^.}- {{\beta}^*}{{\mu}^.}- {{\gamma}^*}{{\mu}^.}) = 0\\ \nonumber \\ 
\label{eqn:einstein5}
e^{-\nu} ( & & -\frac{1}{2}{{\alpha}^{..}} -\frac{1}{2}{{\beta}^{..}}-\frac{1}{2}{{\gamma}^{..}} -\frac{1}{4}{{\alpha}^.}^2-\frac{1}{4}{{\beta}^.}^2-\frac{1}{4}{{\gamma}^.}^2- \nonumber \\ & & \frac{1}{4}{{\alpha}^.}{{\beta}^.} -\frac{1}{4}{{\alpha}^.}{{\gamma}^.} - \frac{1}{4}{{\beta}^.}{{\gamma}^.} + \frac{1}{4}{{\alpha}^.}{{\nu}^.} +\frac{1}{4}{{\beta}^.}{{\nu}^.} + \frac{1}{4}{{\gamma}^.}{{\nu}^.}) + \nonumber \\ e^{-\mu} ( & & \frac{1}{4} {{\alpha}^*}{{\beta}^*} + \frac{1}{4} {{\alpha}^*}{{\gamma}^*} +\frac{1}{4} {{\beta}^*}{{\gamma}^*} +\frac{1}{4} {{\alpha}^*}{{\nu}^*} + \frac{1}{4} {{\beta}^*}{{\nu}^*} + \frac{1}{4}{{\gamma}^*}{{\nu}^*}) = 0
\end{eqnarray}
\section{Properties of the Model}
\label{sec:properties}

We solve these equations (\ref{eqn:einstein0}-\ref{eqn:einstein5}) with (\ref{eqn:density} - \ref{eqn:press3}) using the separation of variables method.  We make the following substitutions:

\begin{eqnarray}
\alpha = 2{p_1}\ln{t}+2{s_1}\ln{\psi}\\
\beta = 2{p_2}\ln{t}+2{s_2}\ln{\psi}\\
\gamma = 2{p_3}\ln{t}+2{s_3}\ln{\psi}\\
\mu = 2{p_4}\ln{t}+2{s_4}\ln{\psi}
\end{eqnarray}
where the $p_i$ and $s_i$ generalize the Kasner parameters.

We note that unlike the $p_i$ parameters, the $s_i$ parameters do not refer to a time evolution, but rather to configurations of the manifold that depend on the fifth coordinate.

Without loss of generality, we rescale the time coordinate by setting $\nu = 0$.

The Einstein equations reduce to the following six relationships:
\begin{eqnarray}
&(&{p_1}{p_2}+{p_1}{p_3}+{p_1}{p_4}+{p_2}{p_3}+{p_2}{p_4}+{p_3}{p_4})(t^{2{p_4}-2}) = \nonumber \\ &(&{s_1}^2+{s_2}^2+{s_3}^2+{s_1}{s_2}+{s_1}{s_3}-{s_1}{s_4}+{s_2}{s_3}-{s_2}{s_4}-{s_3}{s_4}-s_1-s_2-s_3)({\psi}^{-2{s_4}-2})\\ \nonumber \\
&(&{p_2}^2+{p_3}^2+{p_4}^2+{p_2}{p_3}+{p_2}{p_4}+{p_3}{p_4}-p_2-p_3-p_4)(t^{2{p_4}-2}) = \nonumber \\ &(&{s_2}^2+{s_3}^2+{s_2}{s_3}-{s_2}{s_4}-{s_3}{s_4}-s_2-s_3)({\psi}^{-2{s_4}-2})\\ \nonumber \\
&(&{p_1}^2+{p_3}^2+{p_4}^2+{p_1}{p_3}+{p_1}{p_4}+{p_3}{p_4}-p_1-p_3-p_4)(t^{2{p_4}-2}) = \nonumber \\ &(&{s_1}^2+{s_3}^2+{s_1}{s_3}-{s_1}{s_4}-{s_3}{s_4}-s_1-s_3)({\psi}^{-2{s_4}-2})\\ \nonumber \\
&(&{p_1}^2+{p_2}^2+{p_4}^2+{p_1}{p_2}+{p_1}{p_4}+{p_2}{p_4}-p_1-p_2-p_4)(t^{2{p_4}-2}) = \nonumber \\ &(&{s_1}^2+{s_2}^2+{s_1}{s_2}-{s_1}{s_4}-{s_2}{s_4}-s_1-s_2)({\psi}^{-2{s_4}-2})\\ \nonumber \\
&(&{p_1}^2+{p_2}^2+{p_3}^2+{p_1}{p_2}+{p_1}{p_3}+{p_2}{p_3}-p_1-p_2-p_3)(t^{2{p_4}-2}) = \nonumber \\ &(&{s_1}{s_2}+{s_1}{s_3}+{s_2}{s_3})({\psi}^{-2{s_4}-2})\\ \nonumber \\
&&{s_1}{p_1}+{s_2}{p_2}+{s_3}{p_3}-{s_1}{p_4}-{s_2}{p_4}-{s_3}{p_4}=0
\end{eqnarray}

For each equation to hold, its left- and right-hand sides must independently equal zero.  After some manipulation, this yields the following simple relationships for the generalized Kasner parameters:

\begin{eqnarray}
\label{eqn:p1}
&&\sum_{i=1}^{4}{p_i}=1\\
\label{eqn:p2}
&&\sum_{i=1}^{4}{p_i}^2=1\\
\label{eqn:s1}
&&\sum_{j=1}^{3}{s_j}=1+s_4\\
\label{eqn:s2}
&&\sum_{j=1}^{3}{s_j}^2=(1+s_4)^2\\
\label{eqn:ps}
&&\sum_{j=1}^{3}{p_j}{s_j}=p_4 (1+s_4)^2
\end{eqnarray}
Note that (\ref{eqn:p1}-\ref{eqn:ps}) delineate a 3D surface within the 8D parameter space, implying that three of the parameters are independent.  Arbitrarily selecting $p_1$, $p_4$ and $s_4$ as these free parameters, we can express the remaining parameters as:
\begin{eqnarray}
&&p_2 = - \frac{p_1}{2} - \frac{p_4}{2} + \frac{1}{2} \pm \frac{1}{2} \sqrt{-3p_1^2-3p_4^2-2{p_1}{p_4} +2p_1+2p_4+1}\\
&&p_3 = - \frac{p_1}{2} - \frac{p_4}{2} + \frac{1}{2} \mp \frac{1}{2} \sqrt{-3p_1^2-3p_4^2-2{p_1}{p_4} +2p_1+2p_4+1}\\
&&s_1 = \frac{(1+s_4)(p_1+p_4-4{p_1}{p_4}-1+ f)}{4{p_4}^2-2p_4+2}
\end{eqnarray}
where $f=$
\begin{equation}
\sqrt{24{p_4}^4-24{p_4}^3+16{p_1}{p_4}^3+24{p_1}^2{p_4}^2-24{p_1}{p_4}^2-12{p_1}^2{p_4}-3{p_1}^2-3{p_4}^2+6{p_1}{p_4}+2p_1+6p_4+1}
\end{equation}  
\begin{eqnarray}
&&s_2 = -\frac{s_1}{2}+\frac{s_4}{2} +\frac{1}{2} \pm \sqrt{s_4^2+2s_4+1}\\
&&s_3 = -\frac{s_1}{2}+\frac{s_4}{2} +\frac{1}{2} \mp \sqrt{s_4^2+2s_4+1}
\end{eqnarray}

Expression (\ref{eqn:density}) becomes:
\begin{equation}
\rho=\frac{p_4(p_4-1)}{t^2}
\end{equation}
Consequently, in order to guarantee positive induced density, $p_4$ must be negative.  This provides a natural means by which the fifth scale factor contracts over time.

Equations (\ref{eqn:press1}-\ref{eqn:press3}) become:
\begin{eqnarray}
P_1=-\frac{{p_1}{p_4}}{t^2}\\
P_2=-\frac{{p_2}{p_4}}{t^2}\\
P_3=-\frac{{p_3}{p_4}}{t^2}
\end{eqnarray}
Hence the density and pressure are each independent of $\psi$.  In the manner of Socorro, et. al. \cite{socorro}, we can define an effective pressure:
\begin{equation}
P_{eff}= \frac{1}{3} {\sum_{i=1}^3{P_i}} = \frac{p_4(p_4-1)}{3t^2}
\end{equation}
Then we can compose the effective ``equation of state'':
\begin{equation}
\rho=3P_{eff}
\end{equation}
This resembles the equation of state of a hot photon gas, appropriate for the very early universe.  The deviations of the actual induced pressure components from $P_{eff}$ correspond to directional anisotropies in the behavior of these relativistic particles as the universe expands. 

\section{Conclusions and Further Studies}
\label{sec:conclusions}

We have found a class of solutions for the generalized Bianchi type-I cosmology with induced matter.  Resembling the Kasner models, these solutions possess two additional degrees of freedom.  This permits, for instance, all three scale factors to exhibit growth;  whereas in the standard Kasner case, one must contract while the other two expand. 

In general, the scale factors of these models depend not only on time, but also on the higher dimension $\psi$.  The dependence on $\psi$ takes place by means of power laws similar to the temporal behavior.  Note, however, that the $\psi$-dependent terms themselves do not exhibit temporal evolution, but, rather, correspond to various configurations of the manifold.

We have calculated the induced density and pressure associated with these cosmologies, and found simple relationships between them.  As we have found, guaranteeing the positive nature of the induced density naturally leads to dimensional reduction. While compactification is not assumed {\em a priori}, shrinking down of the higher dimension occurs as a consequence of these models' dynamics.

Wesson and Ponce de Leon have speculated that the impact of higher dimensions (via induced matter) could be measured by means of presently observable astrophysical parameters, including effects on the peculiar velocity of galaxies \cite{wesson_deleon}.  Wesson has also suggested that Kaluza-Klein solitons--objects derived from higher dimensional theory with some of the features of black holes (but lacking singularities)--represent possible dark matter candidates \cite{wesson3}.  Perhaps the impact of anisotropy on induced matter in the early universe might manifest itself through similarly measurable astrophysical mechanisms.

In recent years, cosmologies based on superstring and M-theories have generated great interest \cite{lidsey_wands_copeland}.  The discovery that dualities between various superstring models follow from classical symmetries of an encompassing M-theory, with supergravity as its low-energy limit \cite{horava_witten1,horava_witten2}, has led to heightened exploration of the cosmological implications of these approaches.  Because the most prominent candidate models are ten- and eleven-dimensional, one naturally wonders what mechanism confines present-day observed matter to four non-compact dimensions. Compactification represents but one possibility, in which extra-dimensions are curled into one of the Calabi-Yau configurations.  

Randall and Sundrum have recently made an intriguing alternative suggestion in which we reside in a universe of more than four non-compact dimensions\cite{randall_sundrum}.  They examine a 3-brane (3+1 dimensional subspace) in a space of higher dimesions, and show how standard gravitational theory might be reproduced in a low-energy limit without assuming that the higher dimensions are compact.  They demonstrate that for certain tensions of the brane, the probability of losing energy to the higher dimensional modalities would be sufficiently low that known tests of gravity would be upheld.

As this study indicates, dynamical dimensional reduction represents yet another alternative.  Without presupposition of compactness, the higher-dimensional equivalent of general relativistic dynamics might naturally lead to a shrinking of extra scale factors.

This present study has focused on a higher dimensional analogy of Kasner cosmologies.  In four dimensions, Kasner models display the simplest type of anisotropic dynamics, monotonic evolution of the scale factors.  Research has shown that more complex types of anisotropic behavior can be represented by transitions between epochs of Kasner-like behavior.  In particular, the chaotic ``Mixmaster'' type dynamics of Bianchi types VIII and IX can be constructed by means of pasting together the asymptotic behavior of successive Kasner solutions.  Furthermore, as Belinskii, Khalatnikov and Lifshitz proved, never-ending oscillatory behavior represents the general approach to the initial singularity for the full solution of the four-dimensional Einstein vacuum equations \cite{bkl}.

Considering these important results, much work has been done examining the question of chaos in Kaluza-Klein models.  While chaotic oscillatory behavior is absent in all models of dimension $11$ or higher \cite{demaret_henneaux_spindel}, as well as in diagonal models of dimension $5$ or higher, Demaret, de Rop and Henneaux have shown that chaos can be reestablished for space-time dimensions between $5$ and $10$ in the homogeneous vacuum case when off-diagonal terms are included \cite{demaret_derop_henneaux}.

Barrow and D\c{a}browski have investigated the possibility of chaos in string cosmology.  Using a Hamiltonian analysis, they considered Bianchi-type IX models with a low-energy effective action for bosonic string theory.  (This action is of interest because the solutions it generates display a kinetic-energy driven universal dynamics known as pre-big-bang inflation \cite{veneziano,gasperini_veneziano}).  Barrow and D\c{a}browski found that unlike the four-dimensional vacuum case, the universe engages in a only finite number of oscillations, then maintains monotonic Kasner behavior.  They proposed that the need for duality symmetry appears to be incompatible with chaos \cite{barrow_dabrowski}.  

D\c{a}browski has recently examined generalizations of Bianchi-type I and IX models within the context of Ho\v{r}ava-Witten cosmology.  He has investigated the situation in which six of the eleven dimensions of M-theory are compactified on a Calabi-Yau space, leaving five non-compact dimensions:  a homogeneous 3-space, the time coordinate, and a fifth coordinate that is a $S^1/Z_2$ orbifold.  His study has focussed on whether or not chaotic dynamics would be possible during such a five-dimensional era for cosmologies with Bianchi-type IX 3-space geometries.  In exploring the solution space for such models, he has discovered it to be divided into two classes:  a small region of the parameter space, centered on isotropic solutions, for which oscillatory behavior is strictly impossible, and the remainder of the space, for which oscillations from one Kasner epoch to the next can begin.  However, in the latter case, once the oscillations drive the universe into the near-isotropic region, they immediately cease.  Hence, fully chaotic dynamics are impossible for such cosmologies \cite{dabrowski}.

In yet another interesting recent development, Damour and Henneaux have demonstrated that chaotic oscillatory behavior can be restored for some higher dimensional models by the presence of p-forms in the field spectrum of superstring- and M-theories. As these researchers have shown, the question of whether or not oscillatory transitions continue indefinitely depends upon the coupling strength between the p-forms and the dilaton field.  In the absence of p-forms, or for low coupling strengths, monotonic behavior of the Kasner sort remains stable and no chaos is exhibited \cite{damour_henneaux1,damour_henneaux2}.

Considering these intriguing new results, it would be interested to see if Mixmaster chaos is possible within the context of induced matter theory.  In examining the Kasner parameter space of our model, we find that there is a finite region that supports isotropic expansion of the three spatial parameters, namely:
\begin{eqnarray}
&&p_1 = p_2 = p_3 = \frac{1}{2}\\
&&p_4 = - \frac{1}{2}\\
&&s_4 = -2\\
&&-1 \leq s_1,s_2,s_3 \leq \frac{1}{3} 
\end{eqnarray}

Hence, we expect that if Mixmaster oscillations were to propel the universe into such a region, the universe would begin to expand monotonically, and chaotic oscillations would cease.  To determine conclusively whether or not such behavior would indeed take place, our future studies will focus on a full examination of the effects of induced matter on the behavior of Bianchi-types VIII and IX.

\end{document}